\documentclass [12pt] {article}
\usepackage{graphicx}
\usepackage{epsfig}
\begin {document}
\begin{center}

{\Large {\bf PRODUCTION OF SECONDARIES IN SOFT p+Pb COLLISIONS AT LHC 
AND ALICE DATA}} \\
\vspace{.5cm}
C. Merino, C. Pajares, and Yu. M. Shabelski$^{*}$ \\
\vspace{.5cm}
Departamento de F\'\i sica de Part\'\i culas, Facultade de F\'\i sica, \\ 
and Instituto Galego de F\'\i sica de Altas Enerx\'\i as (IGFAE), \\ 
Universidade de Santiago de Compostela, Galicia, Spain \\
E-mail: merino@fpaxp1.usc.es, pajares@fpaxp1.usc.es

\vspace{.2cm}
 
$^{*}$ Permanent address: Petersburg Nuclear Physics Institute, \\
NCR Kurchatov Institute \\
Gatchina, St.Petersburg 188350 Russia \\
E-mail: shabelsk@thd.pnpi.spb.ru
\vskip 0.5 truecm

A b s t r a c t
\end{center}

We calculate the inclusive spectra of secondaries produced in soft (minimum 
bias) p+Pb collisions in the framework of Quark-Gluon String Model at LHC 
energy, and by taking into account the inelastic screening corrections
(percolation effects). The role of these effects is expected to be very large
at very high energies, and they should decrease the spectra more than 1.5 
times in the midrapidity region and increase them about 1.5 times in the
fragmentation region at the energy of LHC. 

\vskip 1cm


\newpage

\section{\bf Introduction}

The detailed investigation of p+Pb interactions at the highest LHC energy of 
$\sqrt{s} =$ 5 TeV (4 TeV proton beam and $1.57\cdot A$ TeV Pb beam) makes 
part of the nearest plans by LHC. The investigation of soft p+Pb interactions 
is very interesting because it can give the final answer to the problem of
inelastic shadow corrections \cite{CKTr,MPS} for inclusive particle
production.

The recent data obtained by the ALICE Collaboration \cite{ALICE} confirm the
existance of these corrections at the LHC energy. The magnitude of these inelastic
shadow corrections corresponds to the resulting contribution of complicate Reggeon
diagrams with multipomeron interaction~\cite{CKTr}.

In the present paper we present the predictions of the Quark-Gluon String 
Model (QGSM)~\cite{KTM,Kaid} for the rapidity and $x_F$ distributions of 
secondaries produced in p+Pb at $\sqrt{s} =$ 5 TeV. The QGSM quantitatively
describes many features of high energy production processes, including 
the inclusive spectra of different secondary hadrons produced in high
energy hadron-nucleon \cite{KaPi,Sh,ACKS,AMPS}, hadron-nucleus 
\cite{KTMS,Sh1}, and nucleus-nucleus \cite{JDDS} collisions. 
The Monte Carlo version of QGSM is described in~\cite{Blei}.
The model parameters used in the present calculations were fixed by comparison 
of the theoretical calculations to the experimental data.

In the case of interaction with a nuclear target the Multiple Scattering
Theory (Gribov-Glauber Theory) is used, what allows to consider the
interaction with the nuclear target as the superposition of interactions
with different numbers of target nucleons. However, it was shown
in~\cite{CKTr,MPS} that the description of the inclusive spectra of secondaries 
produced in d+Au collisions at $\sqrt{s} = 200$~GeV (RHIC) requires to 
account for the inelastic shadow corrections. These corrections are connected
to the multipomeron interactions and they lead to the saturation of the
inclusive density of secondary hadrons in the soft (low $p_T$) region, where
the methods based on perturbative QCD cannot be used. The effects of the 
inelastic shadow corrections should increase with the initial energy.
The difference in the calculations with and without these effects at LHC
energies is of about 3 times in the midrapidity region and of about 2 times
in the fragmentation region.   

Other predictions for secondary production in p+Pb collisions at LHC
energies (mainly for hard collisions can be found 
in~\cite{Tyw,Arm,Arm1,Sal,Pop}.

\section{\bf Inclusive spectra of secondary hadrons in the \newline
Quark-Gluon String Model}

In order to produce quantitative predictions a model for multiparticle
production is needed. For that purpose we have used the QGSM~\cite{KTM,Kaid}
in the numerical calculations presented below.

In the QGSM high energy hadron--nucleon and hadron--nucleus interactions are
considered as proceeding via the exchange of one or several Pomerons, and
all elastic and inelastic processes result from cutting through or between
Pomerons \cite{AGK}. Each Pomeron corresponds to a cylinder diagram (see
Fig.~1a) that, when cutted, produces two showers of secondaries as it is
shown in Fig.~1b. The inclusive spectrum of secondaries is determined by the
convolution of diquark, valence quark, and sea quark distributions, $u(x,n)$,
in the incident particles, with the fragmentation functions, $G(z)$, of quarks
and diquarks into secondary hadrons. Both functions $u(x,n)$ and $G(z)$ are
determined by the corresponding Reggeon diagrams~\cite{Kai}.

The diquark and quark distribution functions depend on the number $n$ of cut 
Pomerons in the considered diagram. 
There exists some freedom on how to share the initial energy among the
Pomerons at $n > 1$ \cite{KTMS,Hla}. In the following calculations we use the
receipt of~\cite{KTMS}.

\begin{figure}[htb]
\centering
\includegraphics[width=.65\hsize]{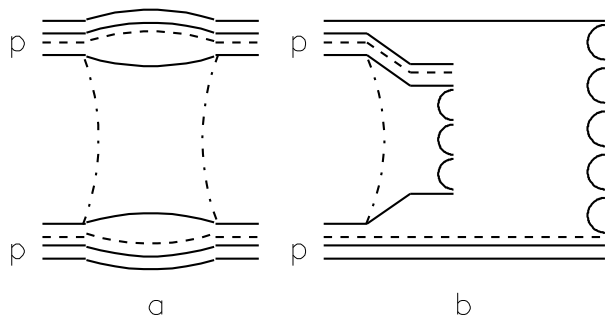}
\vskip -.8cm 
\caption{\footnotesize a) Cylinder diagram (cylinder is shown by
dash-dotted curves) corresponding to the one--Pomeron exchange
contribution to elastic $pp$ scattering, b) and the corresponding
cut diagram which represents its contribution to the inelastic $pp$ cross 
section. Quarks are shown by solid curves and string junctions by
dashed lines.}
\end{figure}

In the case of a nucleon target the inclusive spectrum of a
secondary hadron $h$ has the form \cite{KTM}:

\begin{equation}
\frac{dn}{dy} =  \frac{1}{\sigma_{inel}}\cdot\frac{d\sigma}{dy}
= \frac{x_E}{\sigma_{inel}}\cdot\frac{d\sigma}{dx_F}
=\sum_{n=1}^{\infty}w_{n}\cdot\phi_{n}^{h}(x)\ \ ,
\end{equation}
where the functions $\phi_{n}^{h}(x)$ determine the contribution of diagrams 
with $n$ cut Pomerons and $w_{n}$ is the probability for this process to
occur~\cite{TM}. Here we neglect the diffraction dissociation contributions which 
are important mainly for the secondary production in the large $x_F$ region 
that is not significant in the present calculations.

For $pp$ collisions
\begin{equation}
\phi_n^{h}(x) = f_{qq}^{h}(x_{+},n) \cdot f_{q}^{h}(x_{-},n) +
f_{q}^{h}(x_{+},n) \cdot f_{qq}^{h}(x_{-},n) +
2(n-1)f_{s}^{h}(x_{+},n) \cdot f_{s}^{h}(x_{-},n)\ \  ,
\end{equation}

\begin{equation}
x_{\pm} = \frac{1}{2}[\sqrt{4m_{T}^{2}/s+x^{2}}\pm{x}]\ \ ,
\end{equation}
where $f_{qq}$, $f_{q}$, and $f_{s}$ correspond to the contributions
of diquarks, valence quarks, and sea quarks, respectively.

These contributions are determined by the convolution of the diquark and
quark distributions with the fragmentation functions, e.g.,
\begin{equation}
f_{q}^{h}(x_{+},n) = \int_{x_{+}}^{1}
u_{q}(x_{1},n)\cdot G_{q}^{h}(x_{+}/x_{1}) dx_{1}\ \ .
\end{equation}
The diquark and quark distributions, as well as the fragmentation
functions, are determined by Regge intercepts \cite{Kai}. The numerical 
values of the model parameters were presented in ref.~\cite{Sh}.

The probabilities $w_{n}$ in Eq.~(1) are the ratios of the cross sections
corresponding to $n$ cut Pomerons, $\sigma^{(n)}$, to the 
total non-diffractive inelastic $pp$ cross section, $\sigma_{nd}$~\cite{TM}.

The contribution of multipomeron exchanges in high energy $pp$ interactions 
results in a broad distribution of $w_n$. These distributions at three different
energies $\sqrt{s} = 5$ TeV, 200 GeV, and 20 GeV are presented in the left panel
of Fig.~2.
\begin{figure}
\begin{center}
\vskip -1cm
\mbox{\psfig{file=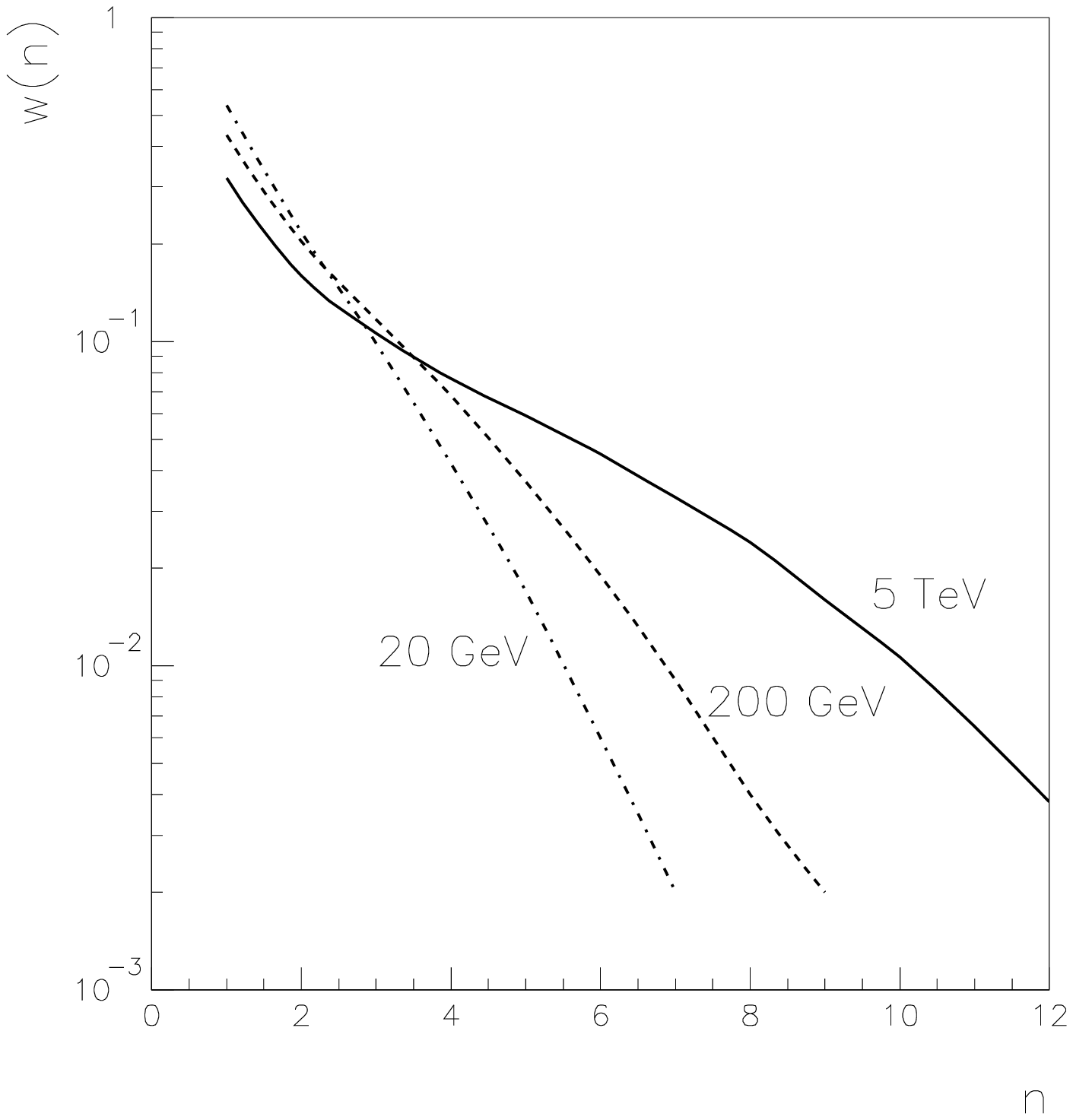,width=0.49\textwidth}}
\mbox{\psfig{file=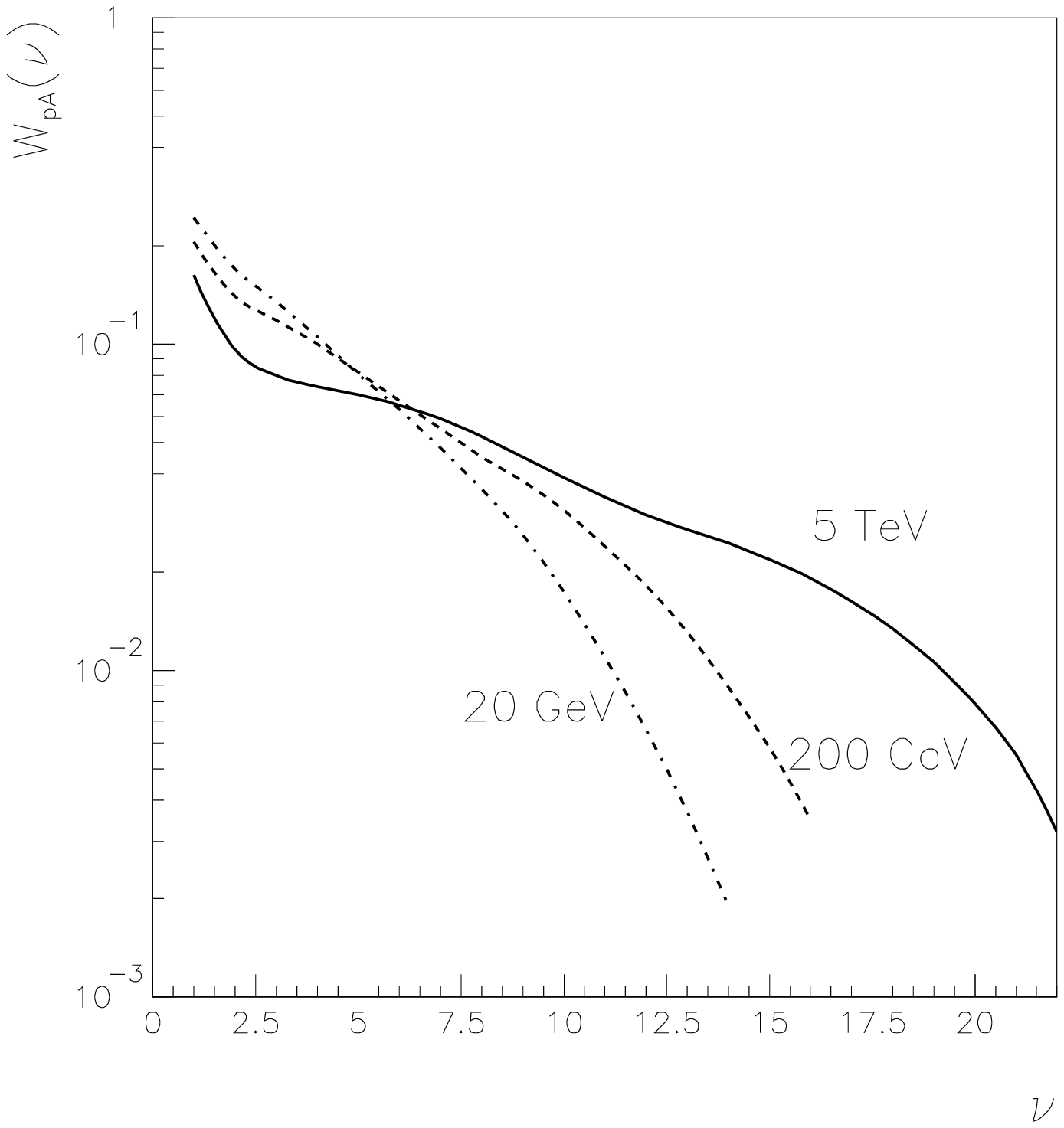,width=0.49\textwidth}}
\vskip -0.5cm
\caption{\footnotesize
The distribution of probabilities to cut $n$ Pomerons, $w_n$ in $pp$
interactions (left), and the distribution of probabilities for the inelastic 
interaction of a proton with $\nu$ lead nucleons, $W_{pPb}(\nu)$, in p+Pb 
interactions (right), at three initial energies, $\sqrt{s} = 5$ TeV 
(solid curves), 200 GeV (dashed curves), and 20 GeV (dash-dotted curves).}
\end{center}
\end{figure}

In the case of nuclear target (or colliding beam) a projectile proton can 
interact with several nucleons. Let $W_{pA}(\nu)$ be the probability for 
the inelastic interactions of the proton with $\nu$ nucleons and
$\sigma_{prod}^{pA}$ the total cross section of secondary production in
a p+A collision. The calculated distributions of $W_{pA}(\nu)$ for proton-lead
interactions at three different energies are presented in the right panel of
Fig.~2. They were calculated in the framework of the Multiple Scattering Theory
as
\begin{equation}
W_{pA}(\nu) = \sigma^{(\nu)}/\sigma_{prod}^{pA} \;,
\end{equation}
where 
\begin{equation}
\sigma^{(\nu)} = \frac1{\nu !} \int d^2b\cdot [\sigma^{pN}_{inel}\cdot T(b)]^{\nu}\cdot
e^{-\sigma^{pN}_{inel}\cdot T(b)}
\end{equation}
coincides~\cite{Sh3,BT,Weis,Jar} with the optical model expression~\cite{TH}, and
\begin{equation}
\sigma_{prod}^{pA} = \int d^2b\cdot(1 - e^{-\sigma^{pN}_{inel}\cdot T(b)}) \:,
\end{equation}
with $T(b)$ being the profile function of the nuclear target:
\begin{equation}
T(b) = A \int^{\infty}_{-\infty} dz\cdot\rho(b,z) \:,
\end{equation}
where $\rho(r=\sqrt{b^2+z^2})$ is the one-particle nuclear density. 

The average value of $\nu$ has the well-known form:
\begin{equation}  
\langle \nu \rangle = \frac{A\cdot\sigma^{pp}_{inel}}{\sigma^{pA}_{prod}} \;.
\end{equation}
We use the values $\sigma^{pp}_{inel} = 72$ mb and $\sigma^{pPb}_{prod} = 1900$ mb
at $\sqrt{s} = 5$ TeV, so the numerical value of $\langle \nu \rangle_{pPb}$ turns out
to be of about 7.9. 

In the calculation of the inclusive spectra of secondaries produced in 
$pA$ collisions we should consider the possibility of one or several Pomeron 
cuts in each of the $\nu$ blobs of proton-nucleon inelastic interactions. 
For example, in Fig.~3 it is shown one of the diagrams contributing to the
inelastic interaction of a beam proton with two lead nucleons. In the 
blob of the proton-nucleon1 interaction one Pomeron is cut, and 
in the blob of the proton-nucleon2 interaction two Pomerons are cut. It is 
essential to take into account all digrams with every possible Pomeron 
configuration and its permutations. The diquark and quark distributions and the 
fragmentation functions here are the same as in the case of the interaction
with one nucleon. 

\begin{figure}[htb]
\centering
\includegraphics[width=.5\hsize]{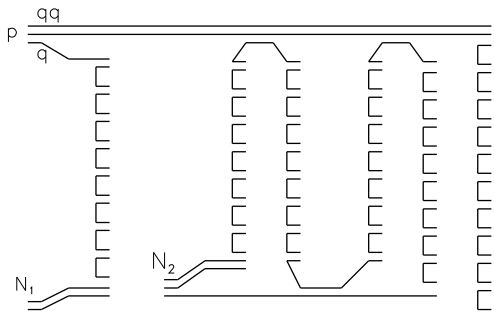}
\caption{\footnotesize One of the diagrams contributing to the inelastic
interaction of an incident proton with two target nucleons $N_1$ and $N_2$
in a $pA$ collision.}
\end{figure}

In particular, the contribution of the diagram in Fig.~3 to the inclusive 
spectrum is
\begin{eqnarray}
\frac{x_E}{\sigma_{prod}^{pA}}\cdot\frac{d \sigma}{dx_F} & = & 2\cdot
W_{pPb}(2)\cdot w^{pN_1}_1\cdot w^{pN_2}_2\cdot\left\{
f^h_{qq}(x_+,3)\cdot f^h_q(x_-,1)\right. + \\ \nonumber & + &
f^h_q(x_+,3)\cdot f^h_{qq}(x_-,1) + f^h_s(x_+,3)\cdot[f^h_{qq}(x_-,2) +
f^h_q(x_-,2) + \\ \nonumber & + & 2\cdot f^h_s(x_-,2)] \left. \right\} \;.
\end{eqnarray}

The process shown in Fig.~3 satisfies~\cite{Sh3,BT,Weis,Jar} the condition
that the absorptive parts of the hadron-nucleus amplitude are determined by
the combination of the absorptive parts of the hadron-nucleon amplitudes.

Sometimes, in the case of hadron-nucleus collisions the values of $x_F$ 
have to be rescaled \cite{Berd,Berd1} to account for the part of
the initial energy that is used for nucleus desintegration, though this
correction becomes negligibly small at LHC energies. 

\section{\bf Inclusive spectra in p+A collisions at very high energy
and inelastic screening (percolation) effects}

The QGSM gives a reasonable description \cite{MPS,KTMS,Sh4} of the inclusive
spectra of different secondaries produced in hadron-nucleus and 
deuteron-nucleus collisions at energies $\sqrt{s_{NN}}$ = 14$-$30 GeV. Also
the experimental data of secondaries produced in Pb+Pb collisions at
$\sqrt{s_{NN}}$ = 17 GeV \cite{NA49,NA49a} were reasonably described 
(except for some problems with charged kaons) by a superposition model 
in~\cite{JDDS}.

At RHIC energies the situation drastically changed. The spectra of 
secondaries produced in $pp$ collisions can be rather well described, but
the RHIC experimental data for Au+Au collisions \cite{Phob,Phen} give clear
evidence of the inclusive density saturation effects which reduce the 
inclusive density about two times in the central (midrapidity) region when 
compared to the predictions based on the superposition picture 
\cite{CMT,Sh6,AP}. This reduction can be explained by the inelastic screening
corrections connected to multipomeron interactions~\cite{CKTr}. The effect is
very small for integrated cross sections (many of them are determined only by
geometry), but it is very important~\cite{CKTr} for the calculations of 
secondary multiplicities and inclusive densities at high energies.

Following the estimations presented in reference~\cite{CKTr}, the RHIC energies
are just of the order of magnitude needed to observe this effect. The inelastic
screening can make \cite{CKTr} the inclusive density to decrease in the 
midrapidity region about two times at RHIC energies and about three times at 
LHC energies, with respect to the calculations without inelastic screening.

However, all estimations are model dependent. The numerical weight of the 
contribution of the multipomeron diagrams is rather unclear due to the many 
unknown vertices in these diagrams. The number of unknown parameters can be 
reduced in some models, and for example in reference~\cite{CKTr} the Schwimmer 
model~\cite{Schw} was used for the numerical estimations.

Another approaches were used in~\cite{Ost}, where the phenomenological 
multipomeron vertices of eikonal type were introduced for enhancement 
diagram summation. In~\cite{Wer} the problem was considered in parton model,
as elastic and inelastic splitting of parton ladders (EPOS model).\footnote{ 
Another (model dependent) possibility to estimate the contribution of the
diagrams with Pomeron interaction comes~\cite{JUR,JUR1,BP,JDDSh,BJP}
from percolation theory. The percolation approach and its previous version, 
the String Fusion Model~\cite{SFM,SFM1,SFM2} predicted the multiplicity
suppression seen at RHIC energies, long before any RHIC data were taken.}

New calculations of inclusive densities and multiplicities in percolation 
theory both in $pp$~\cite{CP1,CP2}, and in heavy ion collisions~\cite{CP2,CP3} are 
in good agreement with the experimental data in a wide energy region.

The percolation model also provides a reasonable description of the 
transverse momentum distribution (at low and intermediate $p_T$) including 
the Cronin effect and the behaviour of the baryon/meson ratio
\cite{Dias,Paj,Paj1}. Most of the effects predicted by percolation can be 
seen as a direct consequence of the strong colour fields produced in the 
collision. This feature is common to other approaches as the Colour Glass 
Condensate~\cite{McL,Ele}, where a $p_T$ scaling is also obtained
\cite{Schaff}. 

In the percolation approach one assumes that if two or several Pomerons 
overlap in transverse space, they fuse in only one Pomeron. When all 
quark-gluon strings (cut Pomerons) are overlapping, the inclusive density 
saturates, reaching its maximal value at a given impact parameter. This 
approach has only one free parameter $\eta$, called the percolation parameter
\begin{equation}
\eta = N_s\cdot\frac{r^2_s}{R^2}\cdot\langle r(y) \rangle \;,
\end{equation}
with $N_s$ the number of produced strings, $r_s$ the string transverse 
radius, and $R$ the radius of the overlapping area. The factor 
$\langle r(y) \rangle$ accounts for the fact that the parton density near 
the ends of the string is smaller that in the central region, where we 
fix $r(0) = 1$. At large rapidities we have $N_s$ strings with differen
parton densities, $r_i(y)$, and
\begin{equation}
N_s\cdot\langle r(y) \rangle = \sum_{i=1}^{N_s} r_i(y) \;.
\end{equation}

As a result, the bare inclusive density $dn/dy \vert_{bare}$ gets reduced,
and we obtain
\begin{equation}
 dn/dy = F(\eta) \cdot dn/dy \vert_{bare} \; ,
\end{equation}
with \cite{BP1}
\begin{equation}
 F(\eta) = \sqrt{\frac{1 - e^{-\eta}}{\eta}} \; .
\end{equation}


In order to account for the percolation effects in the QGSM, it is technically
more simple~\cite{MPS} to consider in the central region the maximal number of 
Pomerons $n_{max}$ emitted by one nucleon. These Pomerons lead, after they are
cutted, to the different final states. Then the contributions of all diagrams with 
$n \leq n_{max}$ are accounted for as at lower energies. The larger number 
of Pomerons $n > n_{max}$ also can be emitted obeying the unitarity constraint,
but due to fusion in the final state (on the quark-gluon string stage) the cut
of $n > n_{max}$ Pomerons results in the same final state as the cut of
$n_{max}$ Pomerons.

By doing this, all model calculations become rather simple and very similar
to the percolation approach. The QGSM fragmentation formalism allows one to 
calculate the integrated over $p_T$ spectra of different secondaries as the
functions of rapidity and $x_F$.


In this scenario we obtain a reasonable agreement with the experimental data 
on the inclusive spectra of secondaries at RHIC energy (see \cite{MPS} with 
$n_{max} = 13$). 

The simplest assumption of this approach corresponding to the general
approach of the Parton Model~\cite{Kan,NNN} is that of neglecting the energy 
dependence of $n_{max}$, i.e. of using a fixed $n_{max} = 13$ at LHC energy.  
However the numerical calculations~\cite{pPbold} result in too strong 
shadow effects, and they are in contradiction with the measurements by the 
ALICE Collaboration~\cite{ALICE}. 

On the other hand, it has been shown in~\cite{JDDCP} that the number of strings
that can be used for the secondary production should increase with the initial
energy. Thus, in the following calculations we use the value $n_{max} = 13$
at the LHC energy $\sqrt{s}$ = 5 TeV, that can be considered as the normalization
to ALICE data for all charged secondaries. The predictive power of our calculation
applies for different sorts of secondaries in midrapidity region, as well as the
calculations in the fragmentation region. 


This scheme seems closer to the point of view of the Parton Model
\cite{Kan,NNN}, the numerical difference between the present secheme and 
\cite{Kan,NNN} coming from the fact that the ratio $r_s^2/R^2$ in Eq.~(11) is 
rather small \cite{MPS}, so the percolation parameter, $\eta$, is also small 
at not very high energies and many $n \leq n_{max}$ independent Pomerons can 
exist before percolation plugs in. That explains why the effects of Pomeron 
(secondary particle) density saturation are small at fixed target energies 
and they become visible only starting from RHIC energies. The similar energy
dependence in the Reggeon calculus \cite{CKTr} is connected to the suppression 
of the diagrams with multipomeron interactions due to the nuclear longitudinal 
form factor.

In the following calculations one additional effect is also accounted for, namely 
the transfer of the baryon charge to large distances in rapidity distances 
through the string junction effect \cite{ACKS,Artr,IOT,RV,Khar,MRS,SJ2}. This 
effect leads to an asymmetry in the production of baryons and antibaryons
in the central region, that is non-zero even at LHC energies \cite{MPRS}. 
In the calculations of these effects we use the following values of the
parameters \cite{MRS}:
\begin{equation}
\alpha_{SJ}\, =\, 0.5\;\; {\rm and} \quad \varepsilon\, =\, 0.0757\,.
\end{equation} 

\section{\bf Rapidity spectra of secondaries at LHC energies}

In the lead-nucleus fragmentation region the contributions of intranuclear 
cascade and/or Fermi-motion of bound nucleons exist. To avoid these 
contributions we consider only the central (midrapidity) and the proton beam 
fragmentation regions.

The results of our calculations for the pseudorapidity density of all charged
secondaries $dn_{ch}/d\eta$ is presented in Fig.~4. The calculations are
normalized to the ALICE experimental point~\cite{ALICE} $dn_{ch}/d\eta = 16.81 \pm 0.71$
measured in the window $\vert \eta_{lab}\vert \le 2$. For comparison we also present
the distribution $dn_{ch}/d\eta$ at the same energy.

\vspace*{-0.1cm}
\begin{figure}[hbt]
\centering
\includegraphics[width=.6\hsize]{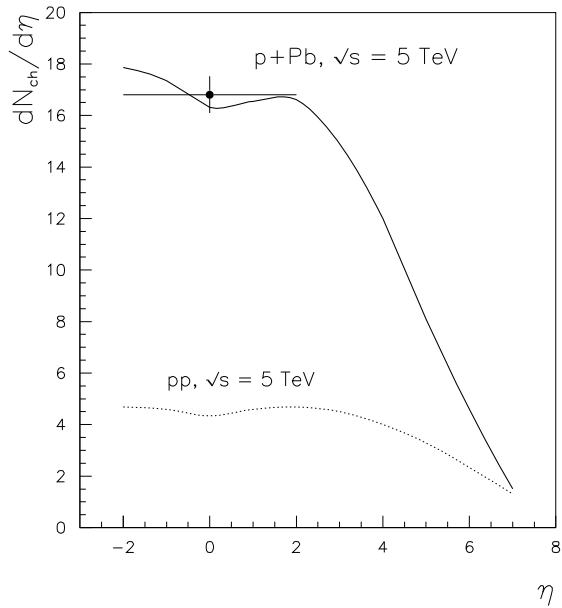}
\vspace*{-0.4cm}
\caption{\footnotesize
Pseudorapidity distributions for all charged secondaries produced in p+Pb 
collisions at $\sqrt{s_{NN}} = 5$ TeV (solid curve) with inelastic screening 
(percolation) corrections, normalized to the ALICE experimental point~\cite{ALICE}.
The dotted curve shows the QGSM predictions for $pp$ collisions at $\sqrt{s_{NN}} = 5$ TeV.}
\end{figure}

It seems more suitable to discuss the spectra behaviour in terms of c.m. 
rapidity $dn/dy_{cm}$.
The rapidity distributions of all charged secondaries
$dn_{ch}/dy_{cm}$, where $y_{cm}$ is determined in the frame of a incident proton interaction 
with one lead nucleon, calculated in the QGSM are presented on the left panel 
of Fig.~5. The solid curve shows the results obtained by accounting for the 
inelastic screening (percolation) corrections, the dashed curve shows the
results obtained without these corrections, and the QGSM predictions for 
$pp$ collisions at $\sqrt{s_{NN}} = 5$ TeV are presented by the dotted curve.

In the right panel of Fig.~5 we present the predictions for the
nuclear modification factor
\begin{equation}
R_{ch}(y_{cm }) = \frac{dn_{ch}/dy_{cm}{\rm (pPb)}}{dn_{ch}/dy_{cm}(pp)}\,.
\end{equation}
Here again, the solid curve shows the QGSM predictions when accounting for
the inelastic screening and the dashed curve corresponds to the results
without these corrections.

\begin{figure}[hbt]
\centering
\includegraphics[width=.49\hsize]{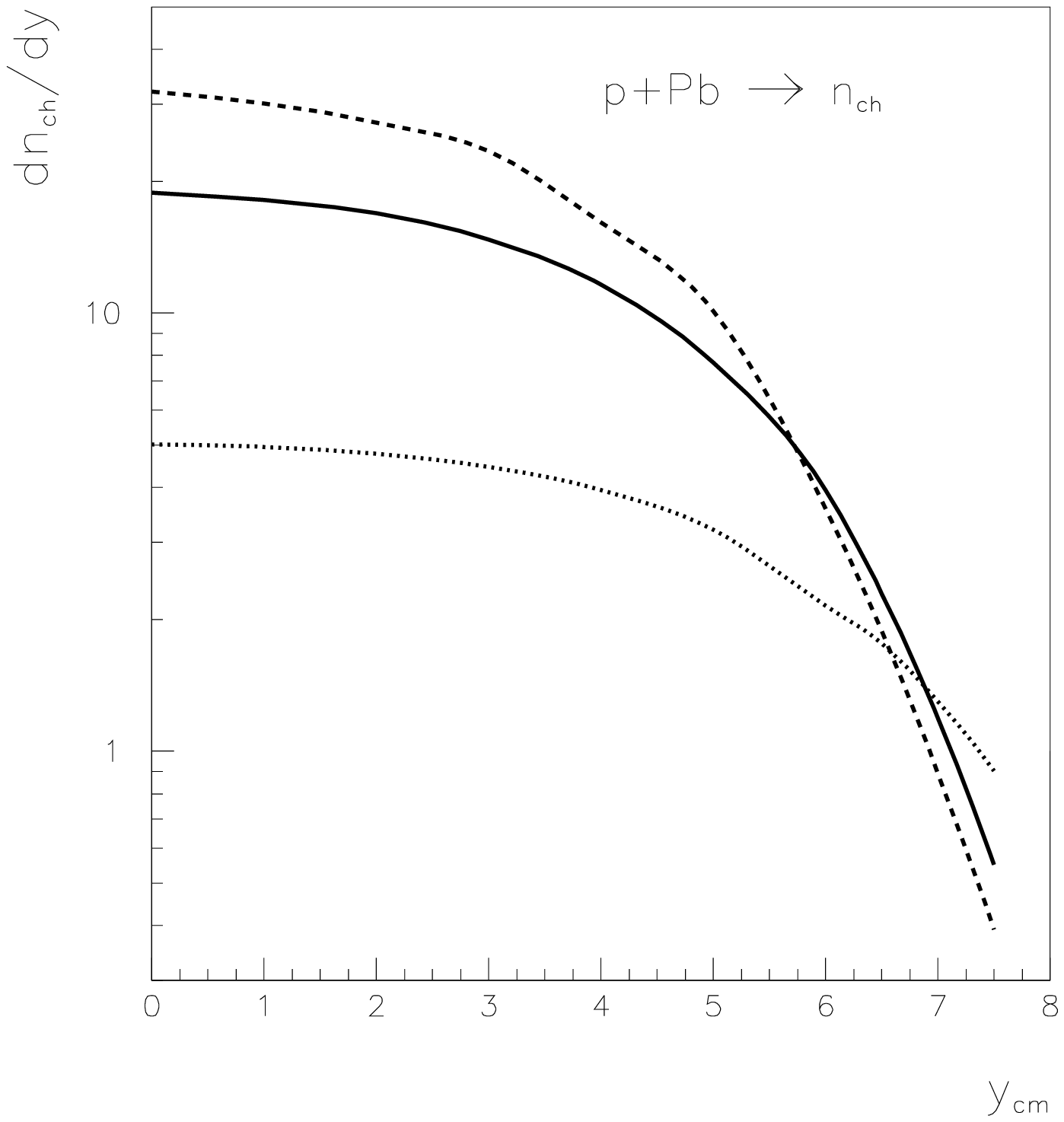}
\includegraphics[width=.49\hsize]{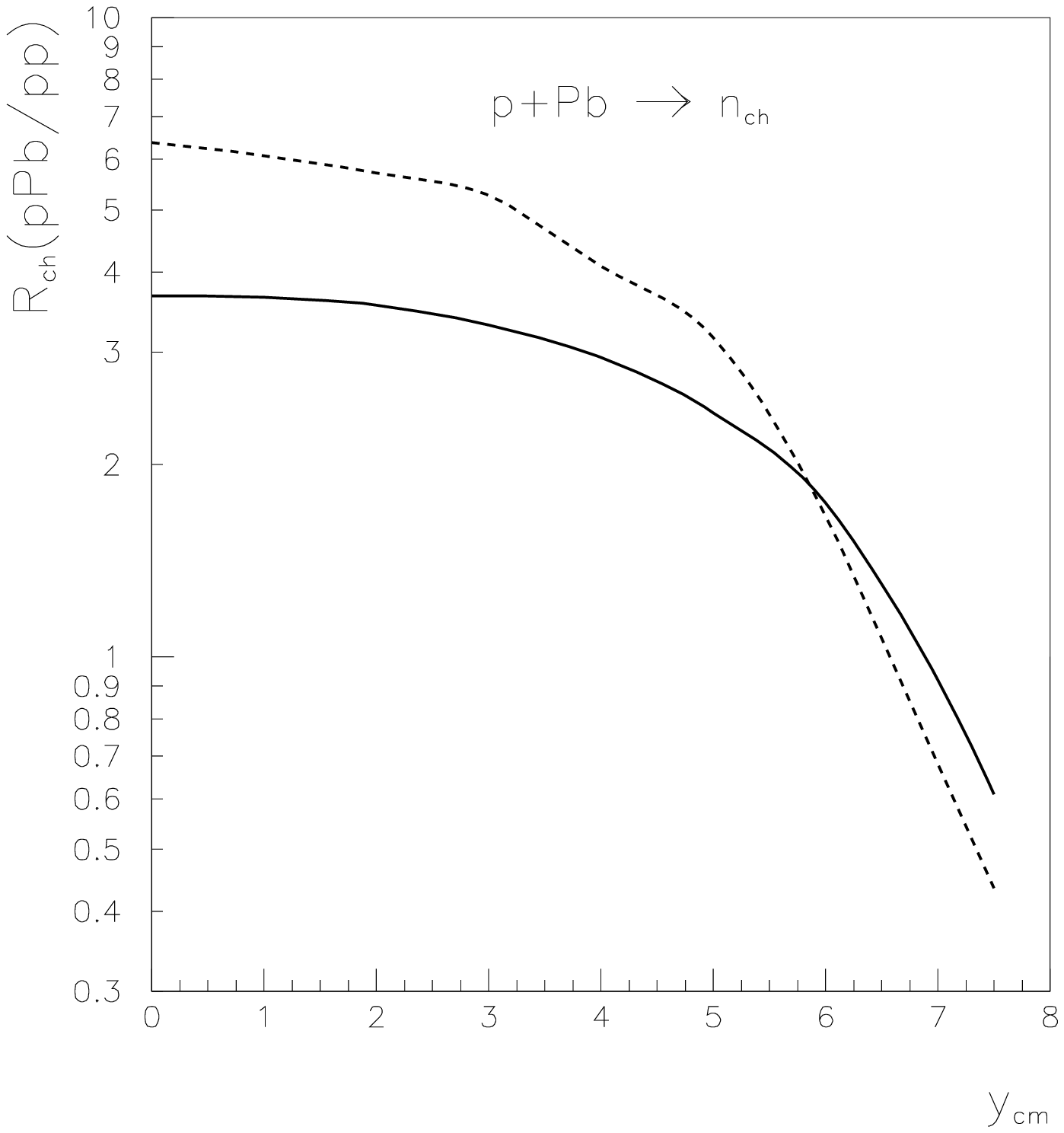}
\vspace*{-0.4cm}
\caption{\footnotesize
Rapidity distributions (left panel) and nuclear modification factors 
(right panel) for all charged secondaries produced in p+Pb collisions 
at $\sqrt{s_{NN}} = 5$ TeV. Dashed curves show the QGSM predictions 
without inelastic screening (percolation) corrections, solid curves show
the QGSM predictions with these corrections, and the dotted curve in the
left panel shows the QGSM predictions for $pp$ collisions at
$\sqrt{s_{NN}} = 5$ TeV.}
\end{figure}

In Fig.~5 one can clearly see that the values of $dn_{ch}/dy$, as well as 
those of the ratios $R_{ch}(pPb/pp)$, differ in the midrapidity region for 
about 1.7 times between the calculations with and without percolation efects, 
what is in agreement with the predictions in reference \cite{CKTr}. These 
differences decrease with rapidity, and at large $y_{cm}$, i.e. in the 
fragmentation region, $y_{c.m.} \geq 6.5$ these differences change sign, the 
values $R_{ch}(pPb/pp)$ becoming smaller than unity in agreement with the 
nuclear shadowing in the inclusive spectra \cite{KTMS,Sh1,ASS}.

The effect of the nuclear shadowing at large rapidities differs for about 
1.5 times for the calculations with and without percolation efects, what is 
a direct consequence of energy conservation. The behaviour of the spectra in 
the proton fragmentation region is discussed in more detail in the next
section.

The results of similar calculations for secondary protons are shown in
Fig.~6. Here again, the difference in the predictions in the midrapidity 
region for the calculations with and without percolation efects is of about 
1.7 times, and these differences decrease with rapidity. The rather 
complicate behaviour of the spectra in the fragmentation region is also 
discussed in the next section.

\vspace{-0.5cm}
\begin{figure}[hbt]
\centering
\includegraphics[width=.49\hsize]{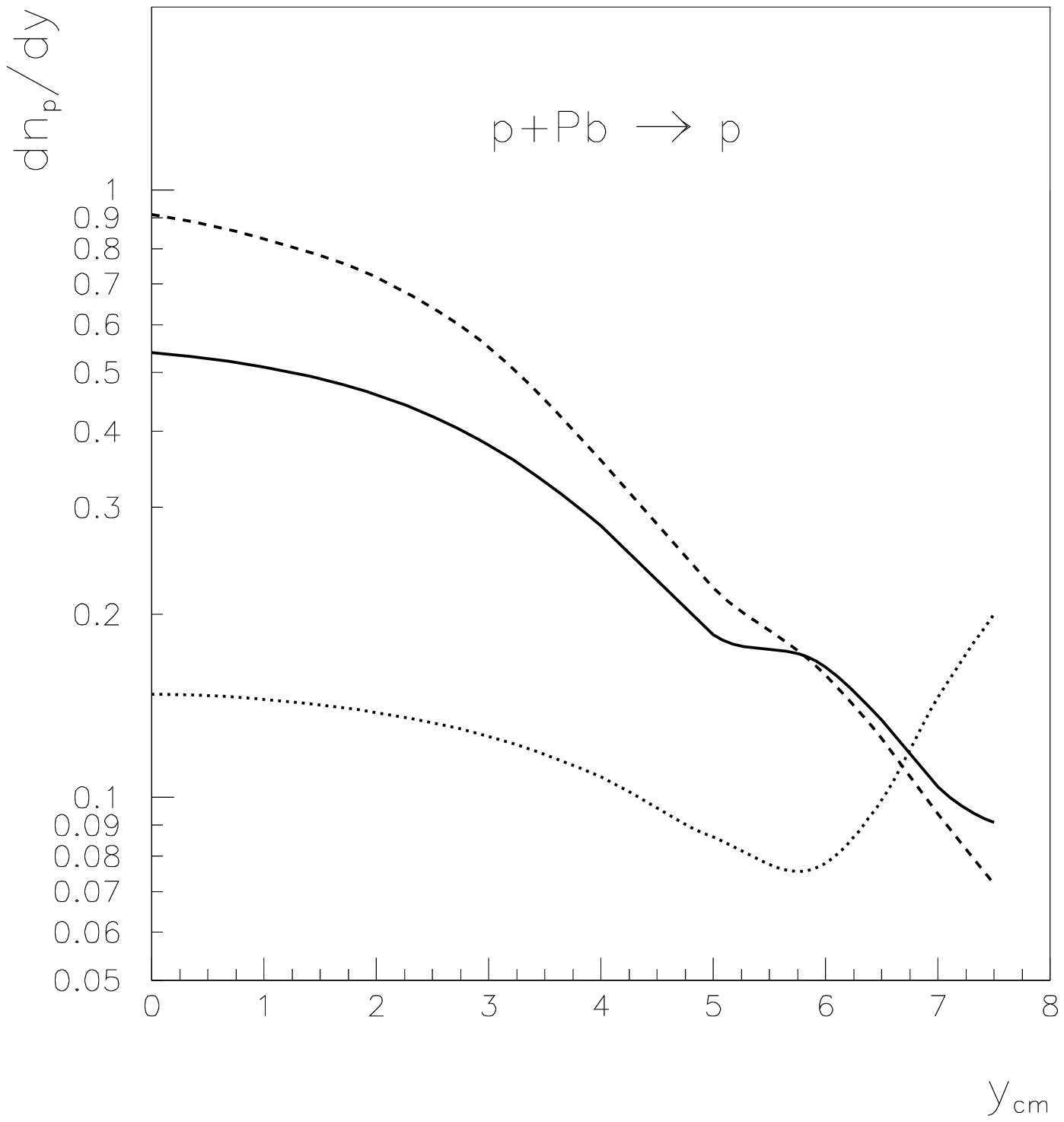}
\includegraphics[width=.49\hsize]{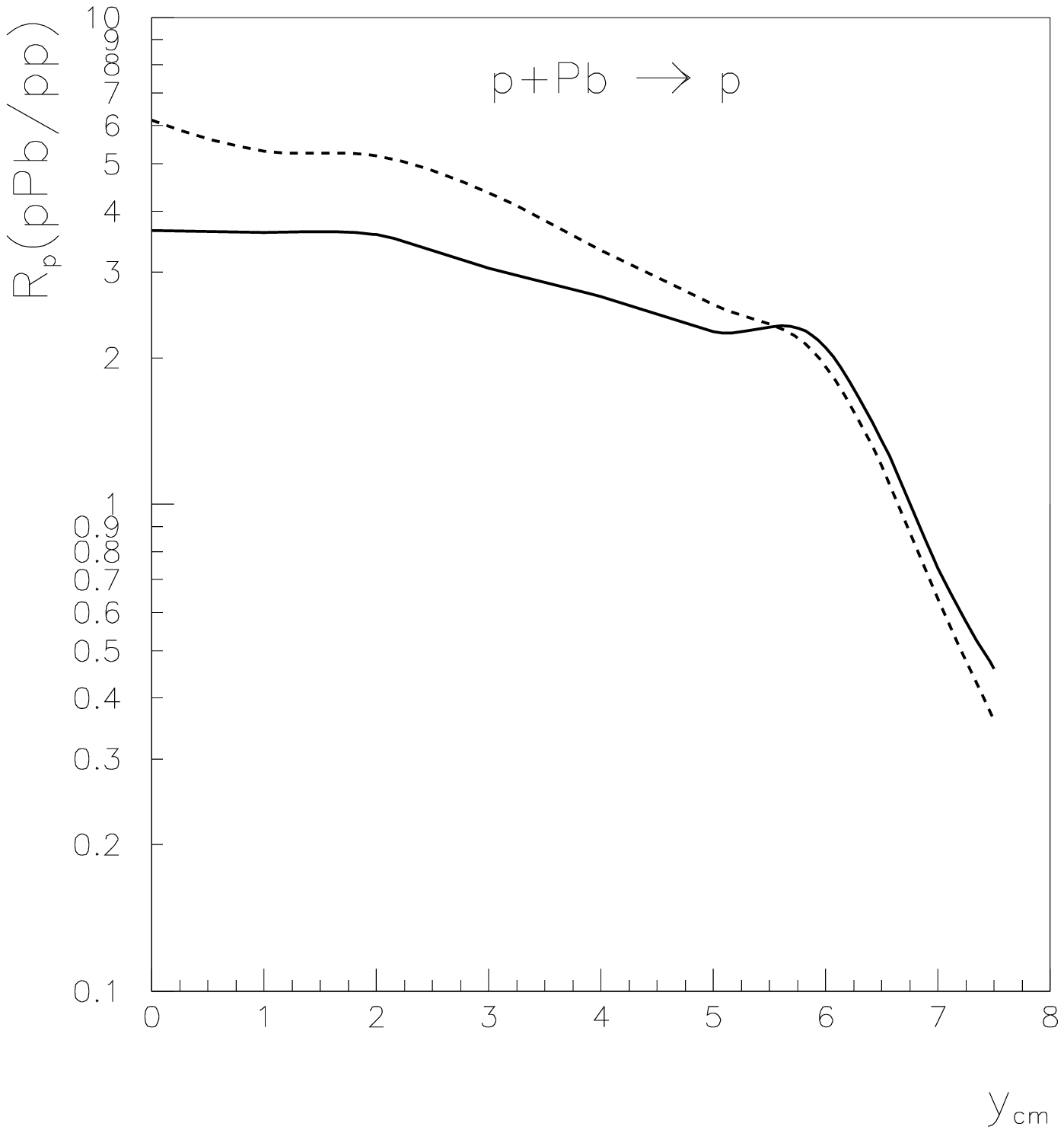}
\caption{\footnotesize
Rapidity distributions (left panel) and nuclear modification factors
(right panel) for secondary protons produced in p+Pb collisions
at $\sqrt{s_{NN}} = 5$ TeV. The correspondence of the curves is the same as
in Fig.~5.}
\end{figure}
\vspace{0.5cm} 

The calculated results of $dn/dy_{cm}$ distributions for secondary $\Lambda$
and $\bar{\Lambda}$ are shown in the upper panels of Fig.~7. The nuclear 
modification factors $R_{\Lambda}(pPb/pp)$ and $R_{\bar{\Lambda}}(pPb/pp)$ 
practically coincide inside the 10$-$15\% accuracy margin, and they are 
presented in the lower panel of Fig.~7.

As a result, we can predict very similar behaviour of the nuclear 
modification factors for different secondaries in the midrapidity region and 
similar dependences of these factors on $y_{cm}$ before the proton
fragmentation regio, say until $R(pPb/pp) \ge 1$. At the same time, these 
factors
are about 1.7 times different at small $y-{c.m.}$ when they are calculated 
with and without inelastic screening corrections.

\vspace{-2.cm}
\begin{figure}[hbt]
\centering
\includegraphics[width=.42\hsize]{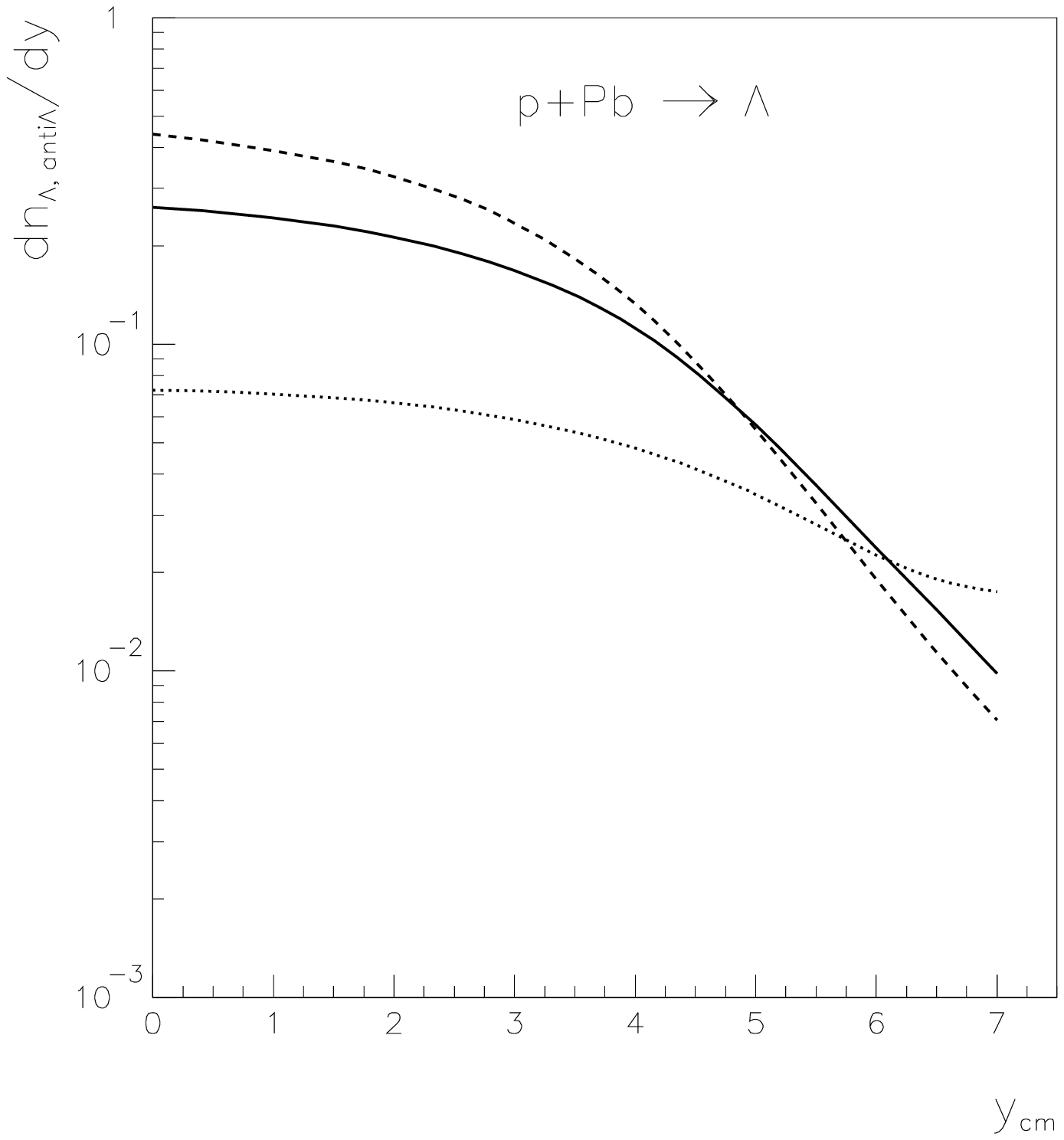}
\includegraphics[width=.42\hsize]{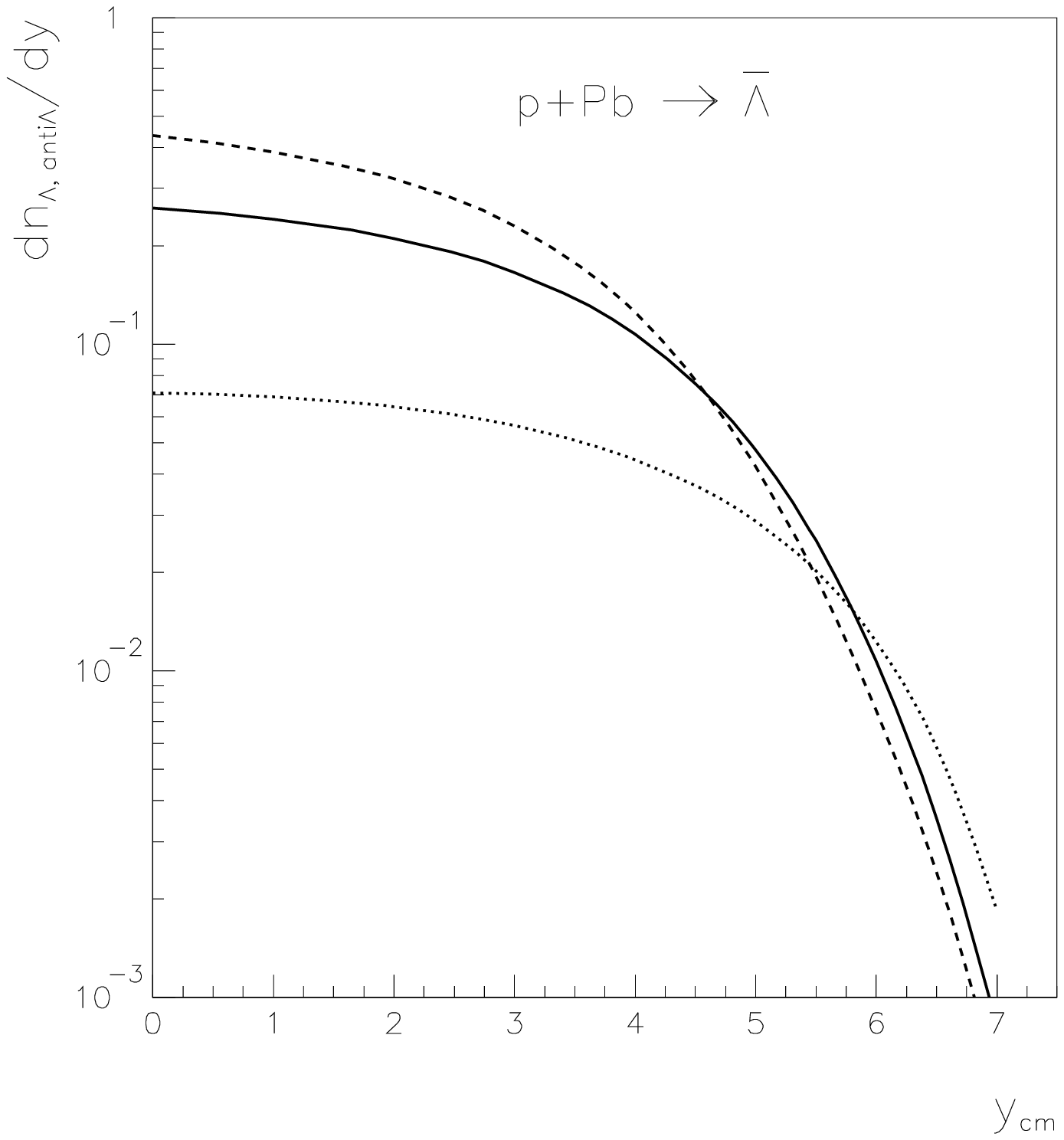}
\includegraphics[width=.42\hsize]{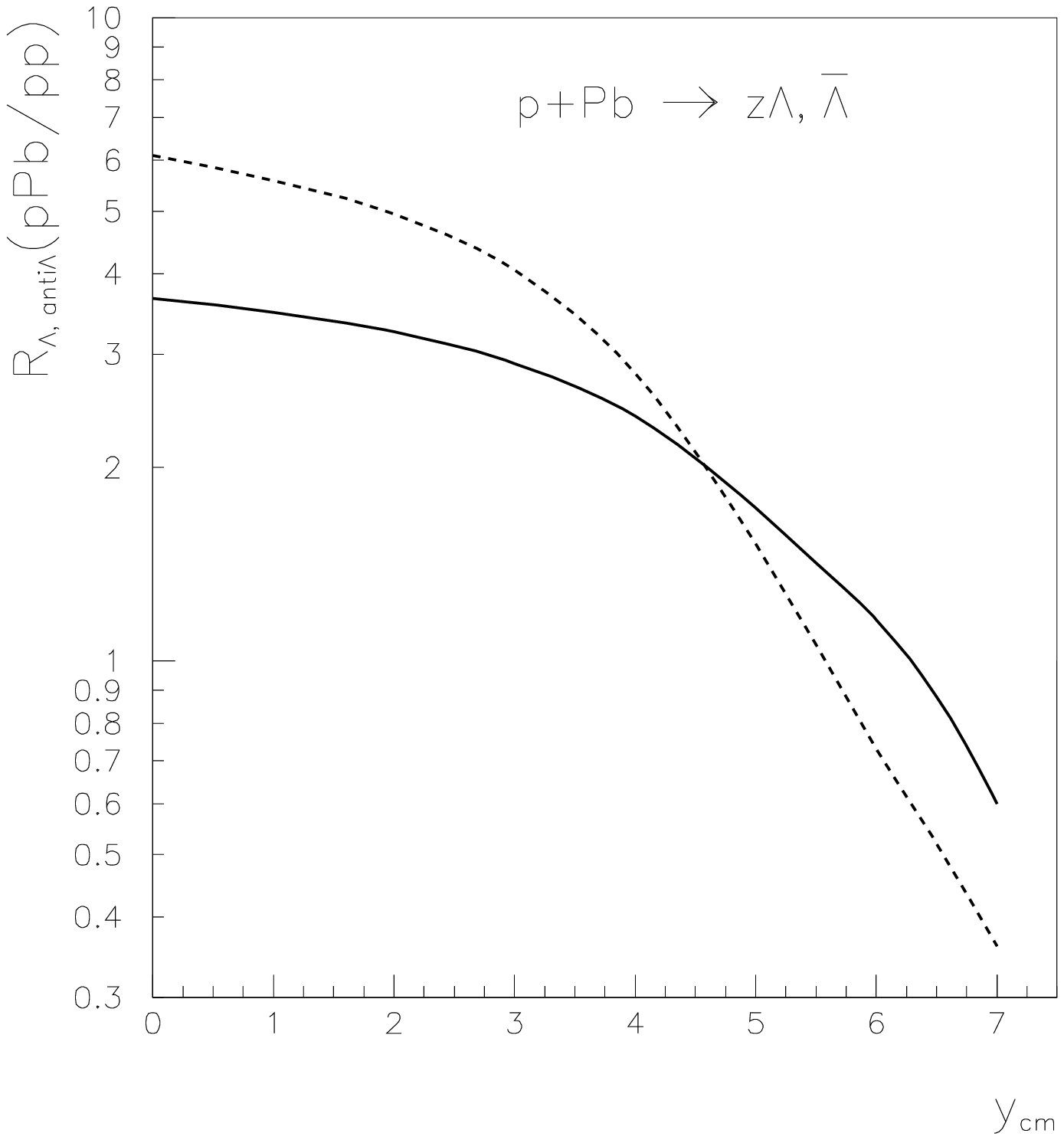}
\vspace{-0.5cm}
\caption{\footnotesize
Rapidity distributions for secondary $\Lambda$ (left up) and
$\bar{\Lambda}$ (right up) produced in p+Pb collisions at
$\sqrt{s_{NN}} = 5$ TeV, and the corresponding nuclear modification
factors for them (low). The correspondence of the curves is the same
as in Fig.~5.}
\end{figure}

\vspace{1.5cm}
\section{\bf Feynman-$x$ spectra at LHC energies}

The $x_F$ variable is mainly suited for the consideration of the inclusive
spectra in the fragnentation region. For the case of the distributions
$(x_E/\sigma_{inel})\cdot(d\sigma/dx_F)$, $x_F$ is determined in the c.m.
frame of the interaction of an incident proton with one lead nucleon.

The distributions $(x_E/\sigma_{inel})\cdot(d\sigma/dx_F)$ calculated in the
QGSM are presented in the left panel of Fig.~8. The solid curve shows the
result obtained with accounting for the inelastic screening (percolation)
corrections, the dashed curve represents the result without these corrections,
and the QGSM predictions for $pp$ collisions at $\sqrt{s_{pp}} = 5$ TeV are
presented by the dotted curve.

In the right panel of Fig.~8 we present the predictions for the nuclear 
modification factor
\begin{equation}
R_{ch}(x_F) = \frac{(1/\sigma_{prod})\cdot(d\sigma_{ch}/dx_F)_{pPb}}
{(1/\sigma_{inel})\cdot(d\sigma_{ch}/dx_F)_{pp}}
\end{equation}
Here again, the solid curve shows the QGSM predictions with accounting for
the inelastic screening and dashed curve shows the result without these
corrections. 

\begin{figure}[hbt]
\centering
\includegraphics[width=.49\hsize]{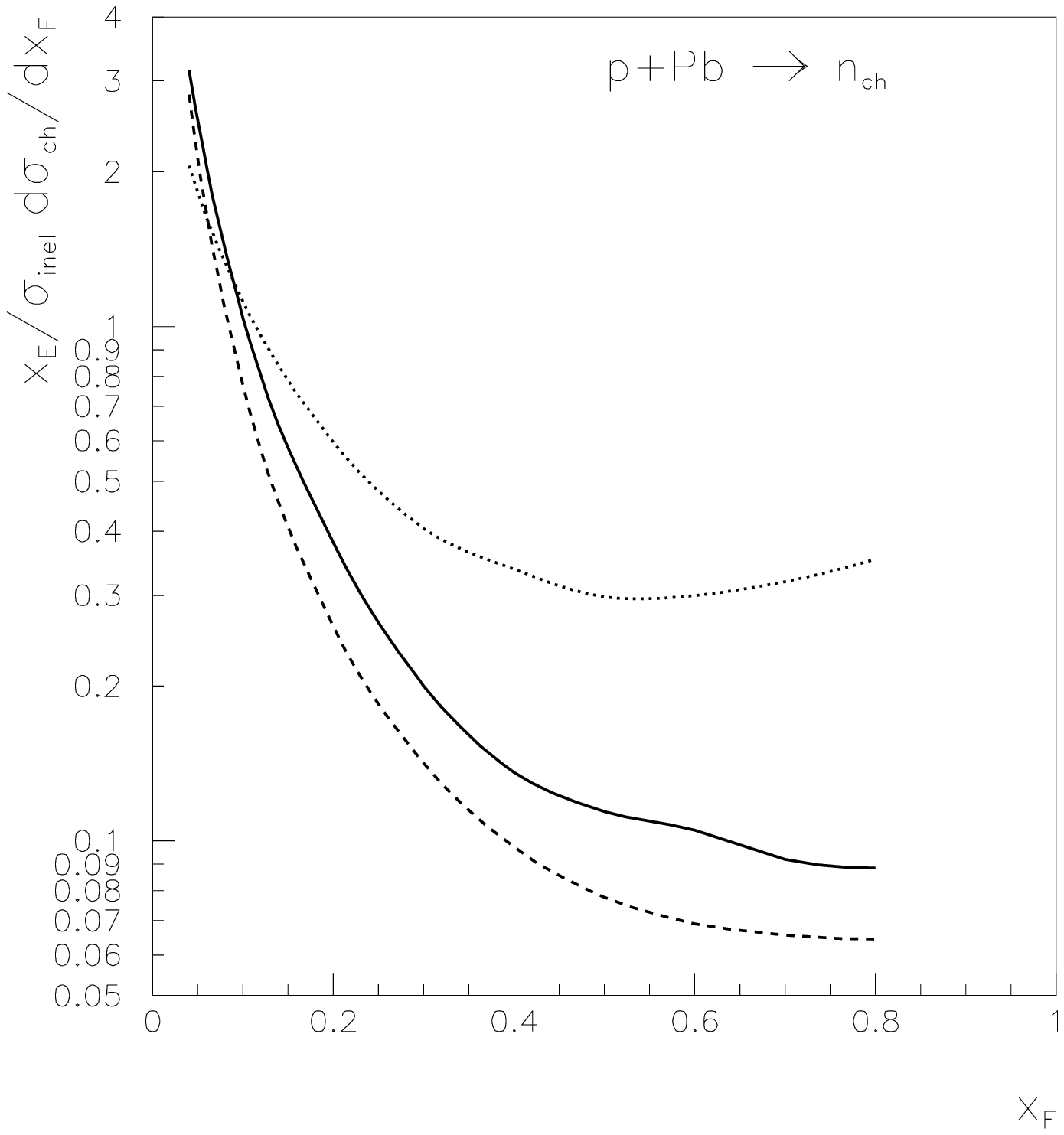}
\includegraphics[width=.49\hsize]{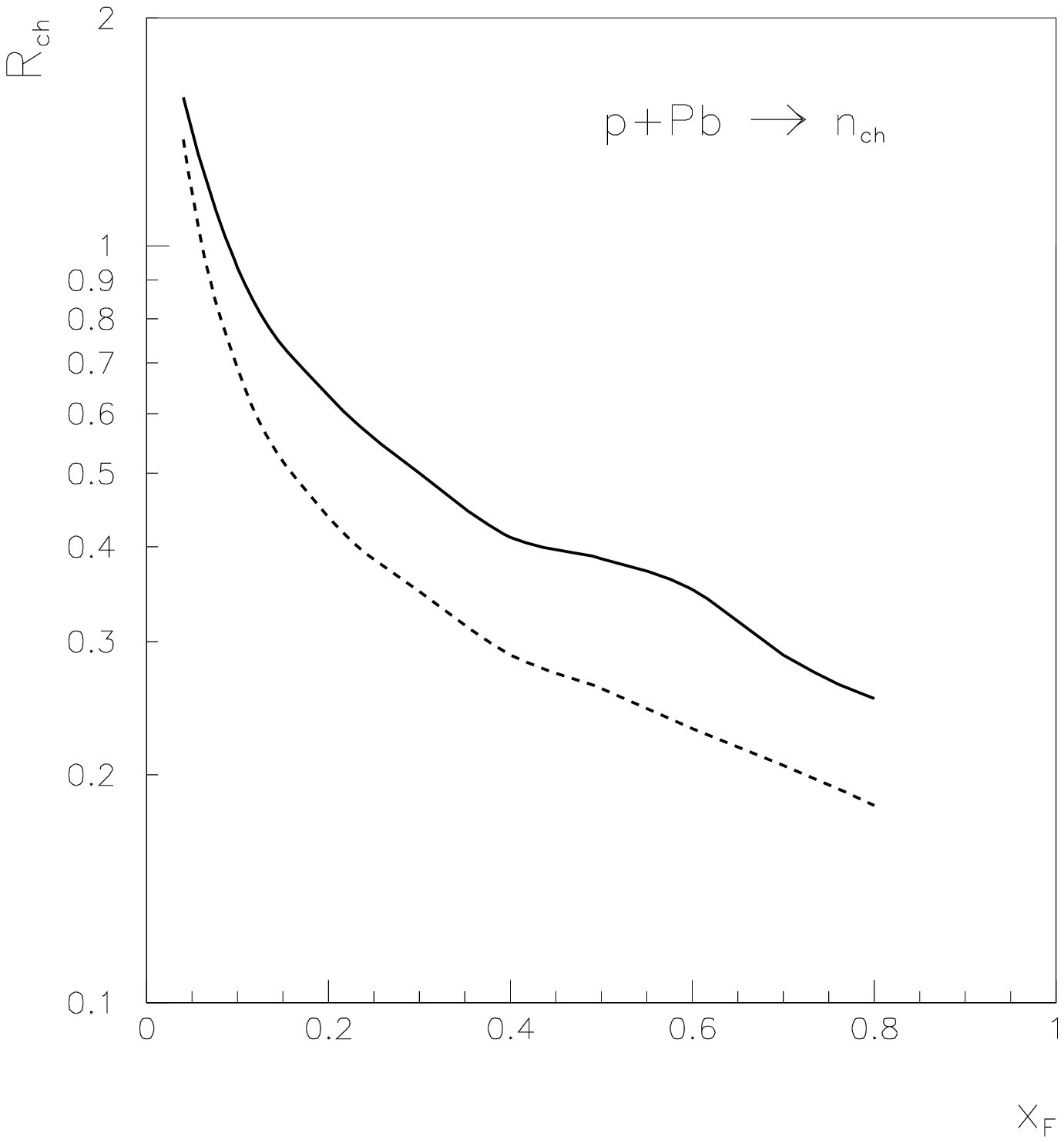}
\vspace*{-0.4cm}
\caption{\footnotesize
Feynman-$x$ distributions (left panel) and nuclear modification factors 
$R_{ch}(x_F)$ (right panel) for all charged secondaries produced in p+Pb 
collisions at $\sqrt{s_{NN}} = 5$ TeV. The correspondence of the curves is
the same as in Fig.~5.}
\end{figure}

The results of the corresponding calculations for the secondary protons are
shown in Fig.~9, where the effects of nuclear shadowing in the fragmentation 
region are very clearly seen, being about 1.5 times different at 
$x_F \geq 1/2$ for the calculations with and without percolation efects.

\begin{figure}[hbt]
\centering
\includegraphics[width=.49\hsize]{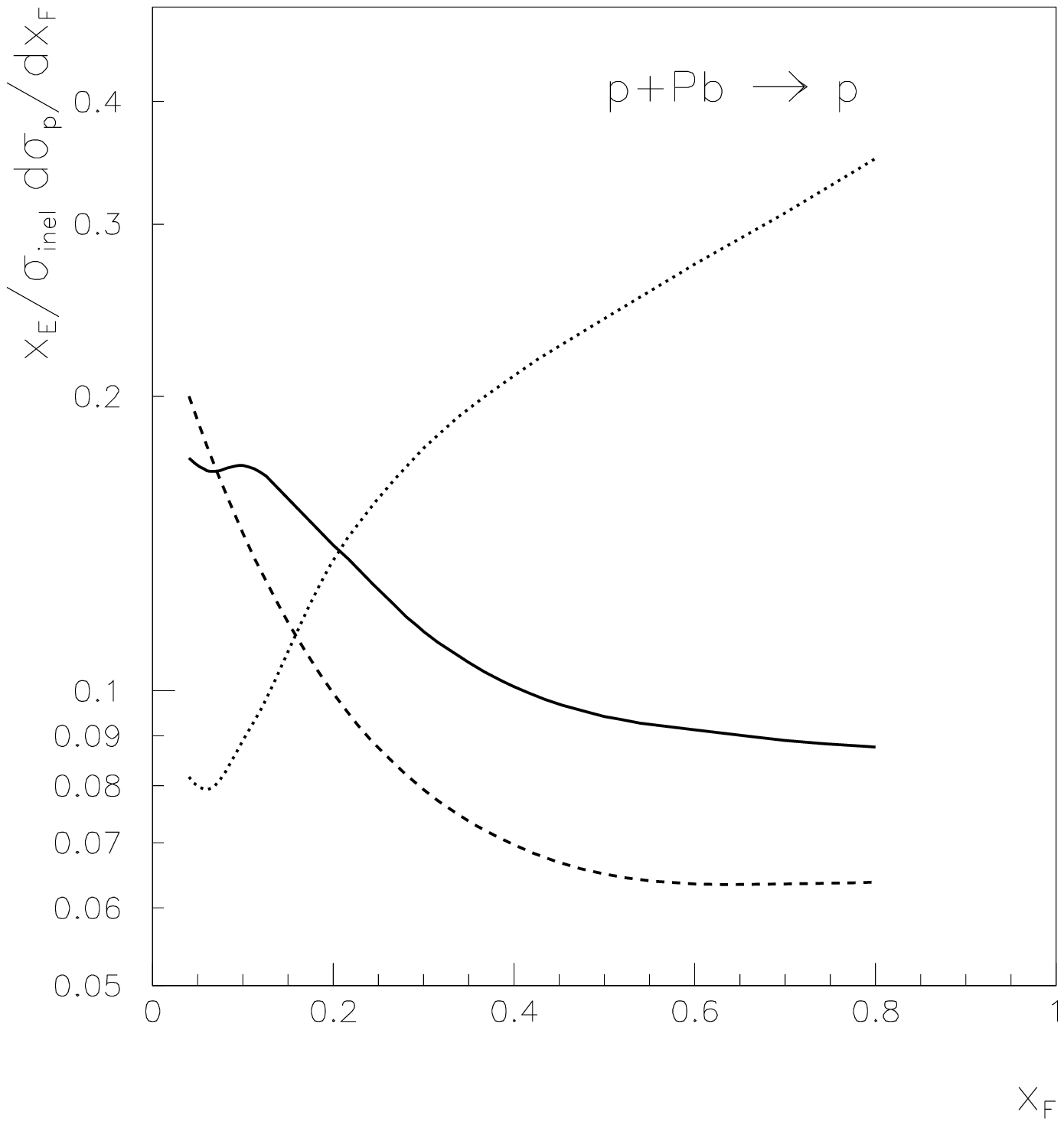}
\includegraphics[width=.49\hsize]{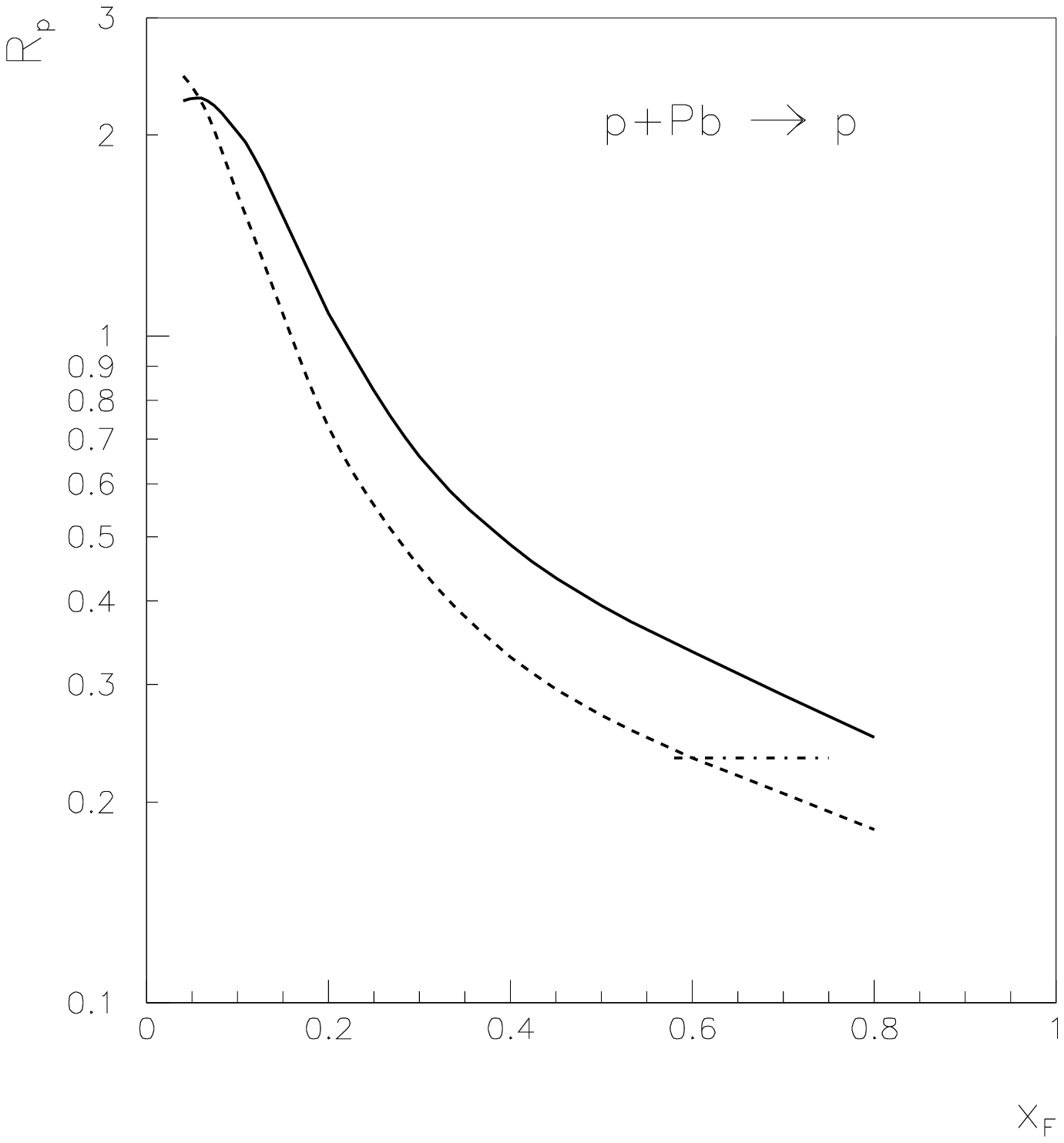}
\vspace*{-0.5cm}
\caption{\footnotesize
Feynman-$x$ distributions (left panel) and nuclear modification factors
(right panel) for secondary protons produced in p+Pb collisions at 
$\sqrt{s_{NN}} = 5$ TeV. The correspondence of the curves is the same as
in Fig.~5. The AQM predictions (eqs.~(19) and (20)) are shown by the 
dash-dotted straight line.}
\end{figure}

Some predictions of the nuclear shadowing in the fragmentation region were
obtained~\cite{ASS} in the framework of the Additive Quark Model (AQM)
In this model the fast baryon is considered as formed by three
constituent quarks, each carrying out around 1/3 of the total baryon momentum 
(sea quarks and gluons are assumed to be effectively included inside the 
constituent quarks). Let a secondary baryon with $x_F \sim 2/3$ to be
formed by two constituent quarks from the initial baryon, the examples being 
the reactions $p \to p$, $p \to \Lambda$, etc. The simplest mechanism in the 
framework of AQM to get this is the recombination of two quark-spectators of
the incident baryon with one newly produced quark from the sea.

The A-dependence of such a processes is determined by the probability to
have two quark-spectators in pA collisions. If the constituent quarks
interact with the target independently, the cross section of their
inelastic interaction with a proton target is
\begin{equation}
\sigma_{inel}^{qN} = \frac13\cdot\sigma_{inel}^{NN}\,,
\end{equation}
and the probability that one constituent quark will interact with the target
nucleus while two other constituent quarks are spectators can be written as:
\begin{equation}
V^{pA}_1 = \frac3{\sigma_{prod}^{pA}}\cdot\int d^2b\cdot 
e^{-2\sigma_{inel}^{qN}\cdot T(b)}
00\cdot\left[1 - e^{-\sigma_{inel}^{qN}\cdot T(b)}\right] \;,
\end{equation}
where $T(b)$ is determined by Eq.~(8).

The AQM predicts an A dependence
\begin{equation}
R_{p,n,\Lambda,...}(x_F \sim 2/3) = V^{pA}_1 \;,
\end{equation}
and such a simple relation is actually in good agreement 
\cite{Sh1,ASS,AS} with the experimental data.

The AQM prediction from Eq.~(20) is shown in the right panel of Fig.~9 by
a dash-dotted straight line, and it is in agreement with the QGSM calculation 
without inelastic shadow corrections. Shadow corrections (elastic as well 
as inelastic ones) also exist in the AQM but they are not accounted for 
in eqs.~(18) and (19).

The calculated results for the $\Lambda$-hyperon production
$(x_E/\sigma_{inel})\cdot(d\sigma_{\Lambda}/dx_F)$ distributions, and the 
corresponding nuclear modification factors $R_{\Lambda}(x_F)$, are shown in 
Fig.~9, where the AQM prediction from Eq.~(20) is presented in the right 
panel of Fig.~10 by the dash-dotted straight line.

\begin{figure}[hbt]
\centering
\includegraphics[width=.49\hsize]{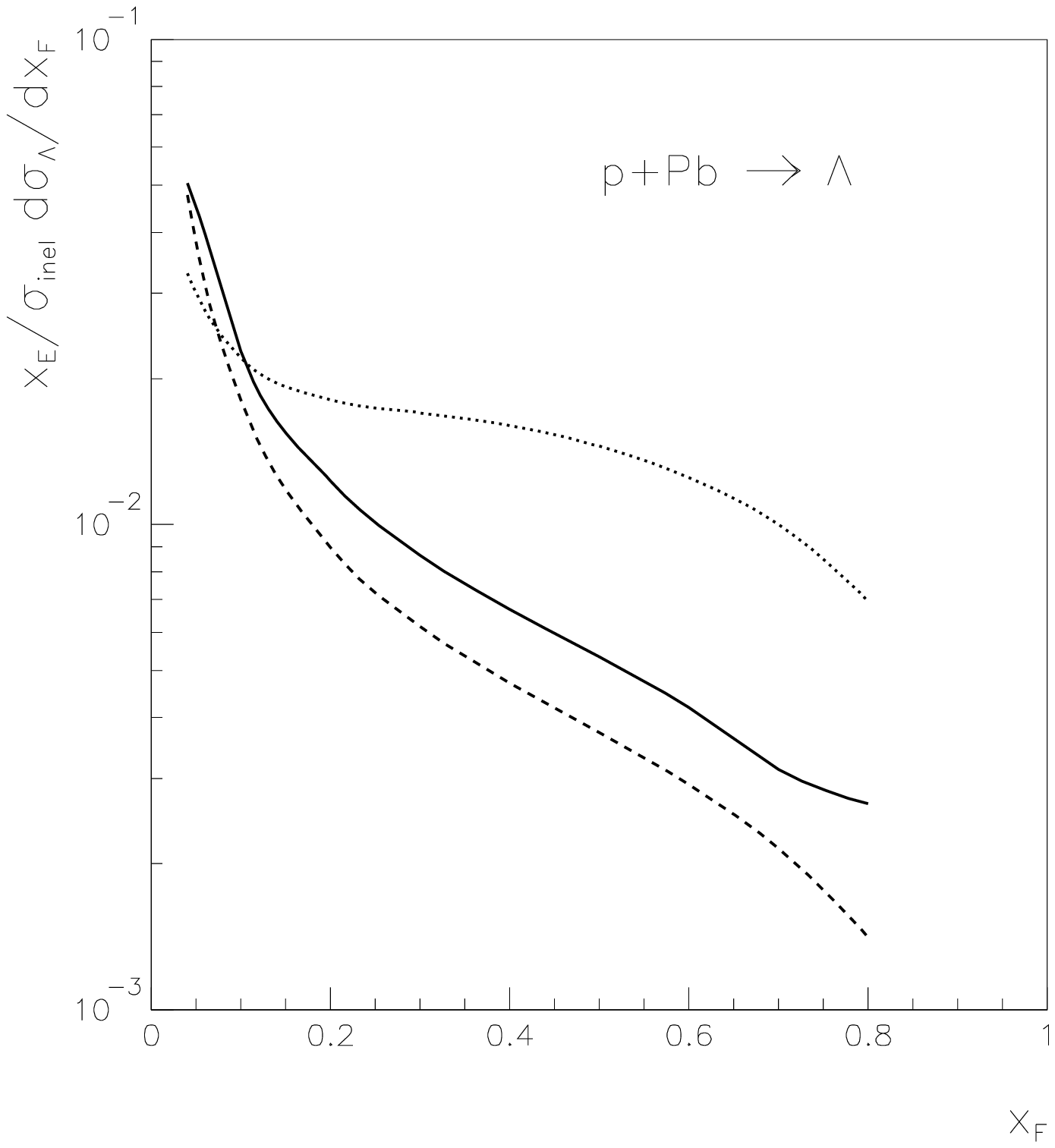}
\includegraphics[width=.49\hsize]{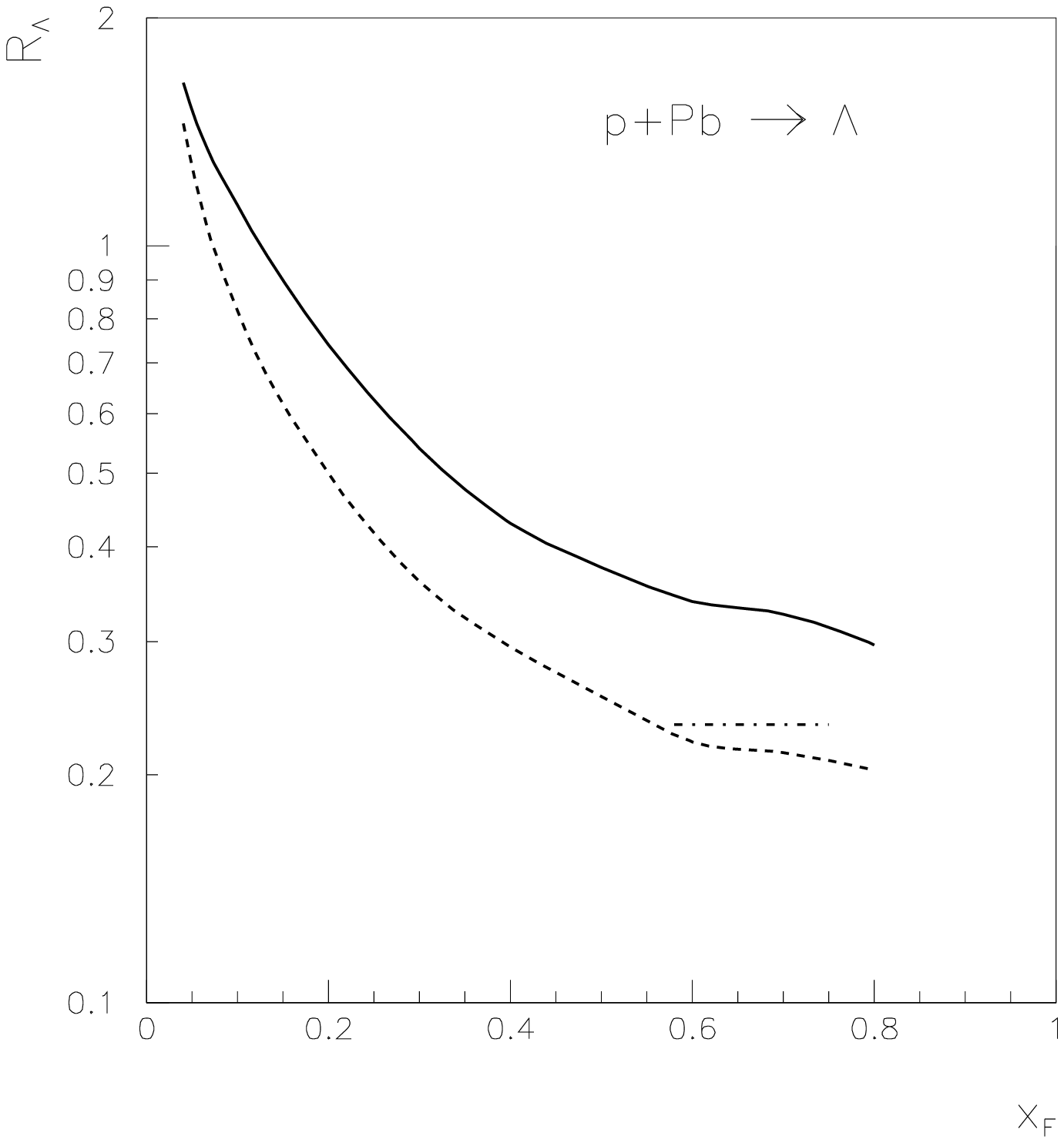}
\vspace*{-0.5cm}
\caption{\footnotesize
Feynman-$x$ distributions (left) and nuclear modification factors
(right) for secondary $\Lambda$ produced in p+Pb collisions at
$\sqrt{s_{NN}} = 5$ TeV. The correspondence of the curves is the same as
in Fig.~5. The AQM predictions from eqs.~(19) and (20) are shown by a
dash-dotted straight line.}
\end{figure}

In Fig.~10 one can clearly appreciate that the spectra of $\bar{\Lambda}$ at
large $x_F$ are very small.

\section{Conclusions}

The percolation (inelastic screening) effects are too small to be observed
by comparing the model calculations to experimental data at fixed target energies 
$\sqrt{s_{NN}} =$ 15$-$40~GeV. At RHIC energies, these effects in
$dn_{ch}/dy (y=0)$ become to be a factor of about 1.5 \cite{MPS}, and they can
increase up to a factor 1.7 at the LHC energy, i.e. a rather slow increase with
energy. A so significant effect at the LHC energies is connected to the fact that at 
$\sqrt{s_{NN}} = 5$ TeV, $\langle n_{NN} \rangle \sim 3$ and 
$\langle \nu_{NPb} \rangle \sim 8$, so the average value of Pomerons 
($\sim 24$) in the minimum bias p+Pb collision is of the same order as 
$n_{max} = 25$, and for larger number of Pomerons (quark-gluon strings) the
fusion processes become very important. These processes decrease the 
inclusive densities (nuclear modification factors $R(y_{cm})$ of Eq.~(16)) 
in the midrapidity region.

Due to the same reason the inclusive densities (the ratios $R(x_F)$ of 
Eq.~(17)) in the fragmentation region should increase. Here we predict the
disagreement with the AQM results.

The detailed experimental confirmation of the inelastic screening (percolation) 
effects for the inclusive spectra can be considered as a natural ``bridge'' 
between the mechanisms of the hadron density saturation in soft and hard 
processes. 

The point to be stressed is the energy dependence~\cite{JDDCP} of $n_{max}$ 
from RHIC to LHC energies, and this dependence should be reproduced when
computing the enhancement Reggeon diagrams.

Our results are in reasonable numerical agreement with those in 
ref.~\cite{DKLN}.

{\bf Acknowledgements}

We are grateful to N. Armesto and J. Dias de Deus for useful discussions. 
This paper was supported by Ministerio Educaci\'on y Ciencia of Spain under 
project FPA 2011-22776, by the Spanish Consolider CPAN Project, and by Xunta
de Galicia (Spain) and, in part, by the grant RFBR 11-02-00120-a.
\vskip 0.1cm


\end {document}